\documentclass[aps,prc,superscriptaddress,nofootinbib,showpacs,floatfix,singlecolumn]{revtex4}
\usepackage{url}
\usepackage{cancel}
\usepackage[colorlinks,linkcolor=blue,citecolor=blue,filecolor=black,urlcolor=blue]{hyperref}
\usepackage{epsfig,graphics}
\usepackage{graphicx}
\usepackage{dcolumn}
\usepackage{bm}
\usepackage[usenames]{color}
\usepackage{amssymb}
\usepackage{amsmath}
\usepackage{multirow}
\usepackage{float}
\usepackage{harpoon}
\usepackage{MnSymbol}
\usepackage{appendix}
\usepackage{color}

\newcommand{\Nch} {N_{\mathrm{ch}}}

\makeatletter

\newcommand{\Rmnum}[1]{\expandafter\@slowromancap\romannumeral #1@}
\makeatother

\makeatletter 
\@addtoreset{equation}{section}
\makeatother  

\newcommand{\sqrtsnn}{\mbox{$\sqrt{s_{\mathrm{NN}}}$}}
\newcommand{\pT} {p_{\mathrm{T}}}
\newcommand{\pPb}{\mbox{$p$+Pb}}
\newcommand{\lr}[1]{\left\langle #1\right\rangle}
\newcommand{\llrr}[1]{\left\llangle #1\right\rrangle}

\newcommand{\nch}{\lr{N_{\mathrm{ch}}}}
\newcommand{\nchb}{\mbox{$N_{\mathrm{ch}}^{\mathrm{sel}}$}}
\newcommand{\scn}{\mathrm{sc}_{n,m}\{4\}}
\newcommand{\sca}{\mathrm{sc}_{2,3}\{4\}}
\newcommand{\scb}{\mathrm{sc}_{2,4}\{4\}}

\newcommand{\acn}[1]{\mathrm{ac}_{#1}\{3\}}

\usepackage{CJK}
\hfuzz=\maxdimen
\begin{document}
\title{Non-flow effects in three-particle mixed-harmonic azimuthal correlations in small collision systems}
\newcommand{\sinap}{Shanghai Institute of Applied Physics, Chinese Academy of Sciences, Shanghai 201800, China}
\newcommand{\cas}{University of Chinese Academy of Sciences, Beijing 100049, China}
\newcommand{\sbu}{Department of Chemistry, Stony Brook University, Stony Brook, NY 11794, USA}
\newcommand{\bnl}{Physics Department, Brookhaven National Laboratory, Upton, NY 11976, USA}
\newcommand{\sari}{Shanghai Advanced Research Institute, Chinese Academy of Sciences, Shanghai 201210, China}
   \author{Chunjian Zhang}\affiliation{\sinap}\affiliation{\cas}\affiliation{\sbu}
   \author{Jiangyong Jia}\email[]{jjia@bnl.gov}\affiliation{\sbu}\affiliation{\bnl}
   \author{Jun Xu}\affiliation{\sari}\affiliation{\sinap}
  \date{\today}
 
\begin{abstract}
The Multi-particle technique has been used to unravel the nature of the long-range collectivity in small collision systems. A large three-particle mixed-harmonic correlation (or event-plane correlation) signal was recently observed by the ATLAS Collaboration, but the role of non-flow correlations is not yet studied. We estimate the influence of non-flow correlations to the three-particle correlators in $pp$ and $p$+Pb collisions using PYTHIA and HIJING models, and compare with the ATLAS results. The large non-flow effects from the jet and dijet production is found to be largely suppressed in $p$+Pb collisions using the subevent cumulant method by calculating the azimuthal correlation between two or more longitudinal pseudorapidity ranges. Depending on the experimental subevent method, however, the non-flow effects may still be significant in $pp$ collisions. Future prospect to measure other multi-particle mixed-harmonic correlators in small systems are discussed.
\end{abstract}

\maketitle

\section{Introduction}
\label{introduction}
In high-energy hadronic collisions, particle correlations are an important tool to study the multi-parton dynamics of QCD in the strongly coupled non-perturbative regime~\cite{Shuryak:2014zxa}. Measurements of azimuthal correlations in small collision systems, such as $pp$ and $p$+A collisions~\cite{CMS:2012qk,Abelev:2012ola,Aad:2012gla,Aad:2014lta,Khachatryan:2015waa}, have revealed a strong harmonic modulation of particle densities d$N/{\textrm d}\phi\propto 1+2\sum_{n=1}^{\infty}v_{n}\cos n(\phi-\Phi_{n})$, where $v_n$ and $\Phi_n$ represent the magnitude and the event-plane angle of the $n^{\mathrm{th}}$-order flow harmonic. They are also conveniently represented by the flow vector $ {\bm V}_n = v_n e^{{\textrm i}n\Phi_n}$. Measurement of ${\bm V}_n$ and their event-by-event fluctuations have been performed as a function of charged particle multiplicity $\Nch$ in $pp$ and $p$+A collisions. It is found that the azimuthal correlations actually involve all particles over a wide pseudorapidity range, similar to those observed in A+A collisions. A key question is whether this multi-particle collectivity reflects initial momentum correlation from gluon saturation effects~\cite{Dusling:2013qoz}, or a final-state hydrodynamic response to the initial transverse collision geometry~\cite{Bozek:2013uha}.

One main challenge in the study of azimuthal correlations in small collision systems is how to distinguish the long-range signal from ``non-flow'' correlations involving only a few particles, mainly from resonance-decays/jets and dijets. These non-flow correlations usually involve particles from one or two localized pseudorapidity regions, and can be reduced by requiring correlation between particles from two or more subevents separated in pseudorapidity. This so-called subevent cumulant method~\cite{Jia:2017hbm} has been validated for correlators involving only the magnitude of the flow harmonics, such as four-particle cumulants $c_n\{4\}=\lr{v_n^4}-2\lr{v_n^2}^2$~\cite{DiFrancesco:2016srj,Jia:2017hbm,Huo:2017nms,Aaboud:2017blb} and four-particle symmetric cumulants $\scn=\lr{v_n^2v_m^2}-\lr{v_n^2}\lr{v_m^2}$~\cite{Huo:2017nms,Aaboud:2018syf}. It is found that $c_n\{4\}$ and $\scn$ from the standard cumulant method are contaminated by non-flow correlations over the full $\Nch$ range in $pp$ collisions and the low $\Nch$ region in $p$+A collisions, while such non-flow correlations are largely suppressed in the subevent method that requires three or more subevents~\cite{Huo:2017nms,Aaboud:2016yar,Nie:2018xog}.

Recently, it is realized that multi-particle event-plane correlators or asymmetric cumulants, involving both the $v_n$ and $\Phi_n$ of the flow vectors, which has been studied extensively in A+A systems~\cite{Aad:2014fla}, can also be used to study the nature of the long-range correlation in small collision systems~\cite{Aaboud:2018syf}. The simplest form of such correlators, i.e., the three-particle asymmetric cumulant $\acn{2,2|4}=\lr{{\bm V}_2^2{\bm V}_{4}^*}=\lr{v_2^2v_{4}\cos4(\Phi_2-\Phi_{4})}$, has been measured by the ATLAS Collaboration~\cite{Aaboud:2018syf}. The advantage of using $\acn{2,2|4}$ is that it is a three-particle correlator, and therefore one can still apply the three-subevent method to suppress the non-flow from dijets. Furthermore, the signal of $\acn{2,2|4}$ scales as $\lr{v_2^2v_{4}}\approx \lr{v_2^4}$ and therefore is comparable to $c_2\{4\}$ and is much larger than $\sca$ and $\scb$. In fact, $\acn{2,2|4}$ is a factor of $1/v_4\sim50$ larger than $\scb$ in $pp$ or $p$+Pb collisions, making it a superior observable to study the multi-particle nature of collectivity in small systems.

The ATLAS results~\cite{Aaboud:2018syf} on $\acn{2,2|4}$ show a clear decrease from the standard to the two-subevent and then the three-subevent methods, which has been interpreted as a systematic suppression of the non-flow correlations.  In this paper, we show explicitly via model simulations that this hierarchy is indeed due to a systematic suppression of the non-flow correlations. We also extend the study to $\acn{2,3|5}=\lr{{\bm V}_2{\bm V}_3{\bm V}_{5}^*}$, which is the next event-plane correlator that could be measured in experiments.

\section{Three-particle asymmetric cumulants and model setup}\label{sec:2}
The framework for the standard cumulant and subevent cumulants are described in Ref.~\cite{Bilandzic:2010jr} and Refs.~\cite{Jia:2017hbm,Aaboud:2018syf}, respectively. The three-particle asymmetric cumulants $\acn{n,m|n+m}$ are obtained from three-particle azimuthal correlations for flow harmonics of order $n$, $m$, and $n+m$ as:
\begin{eqnarray}\label{eq:1}
\acn{n,m|n+m}= \llrr{3}_{n,m|n+m}\;, \lr{3}_{n,m|n+m}=\lr{\mathrm{e}^{{\rm i}(n\phi_1+m\phi_2-(n+m)\phi_3)}}\;,
\end{eqnarray}
One firstly averages all distinct triplets in one event to obtain $\lr{3}_{n,m|n+m}$, then averages over an event ensemble to obtain $\acn{n,m|n+m}$. In the absence of non-flow correlations, $\acn{n,m|n+m}$ measures the correlation between three flow vectors:
\begin{eqnarray}\label{eq:2}
\acn{n,m|n+m}_{\mathrm{flow}}=\lr{{\bm V}_n{\bm V}_m{\bm V}_{n+m}^*}=\lr{v_nv_mv_{n+m}\cos(n\Phi_n+m\Phi_m-(n+m)\Phi_{n+m})}.
\end{eqnarray}

In the standard cumulant method, all triplets are selected using the entire detector acceptance. To suppress the non-flow correlations that typically involve particles emitted within a localized region in pseudorapidity, the particles can be grouped into several subevents, each covering a non-overlapping pseudorapidity interval. The multi-particle correlations are then constructed by correlating particles between different subevents, further reducing non-flow correlations. 

Specifically, in the two-subevent cumulant method, the entire event is divided into two subevents, labeled as $a$ and $b$, for example according to $-\eta_{\rm{max}}<\eta_a<0$ and $0<\eta_b<\eta_{\rm{max}}$. The cumulant is defined by considering all triplets comprised of two particles from one subevent and one particle from the other subevent:
\begin{eqnarray}\label{eq:3}
\acn{n,m|n+m}_{\mathrm{2-sub}} = \llrr{\mathrm{e}^{{\rm i}(n\phi_1^a+m\phi_2^a-(n+m)\phi_3^b)}}.
\end{eqnarray}
where the superscript $a$ ($b$) indicates particles chosen from the subevent $a$ ($b$). The two-subevent method suppresses correlations within a single jet (intra-jet correlations), since each jet usually emits particles to one subevent.

Similarly for the three-subevent method, the $|\eta|<\eta_{\rm{max}}$ range is divided into three equal ranges, and they are labelled as $a$, $b$ and $c$, respectively. The corresponding cumulant is defined as:
\begin{eqnarray}\label{eq:4}
\acn{n,m|n+m}_{\mathrm{3-sub}} = \llrr{\mathrm{e}^{{\rm i}(n\phi_1^a+m\phi_2^b-(n+m)\phi_3^c)}}.
\end{eqnarray}
Since the two jets in a dijet event usually produce particles in at most two subevents, the three-subevent method further suppresses inter-jet correlations associated with dijets. To enhance the statistical precision, the pseudorapidity range for subevent $a$ is also interchanged with that for subevent $b$ and $c$, and these different configurations are averaged to obtain the final result.

To evaluate the influence of non-flow effects to $\acn{n,m|n+m}$ in the standard and subevent method, the PYTHIA8~\cite{Sjostrand:2007gs} and HIJING~\cite{Gyulassy:1994ew} models are used to generate $pp$ events at $\sqrt{s} = 13$ GeV and $\pPb$ events at $\sqrtsnn = 5.02$ TeV, respectively. These models contain significant non-flow correlations from jets, dijets, and resonance decays, which are reasonably tuned to describe the data, such as $\pT$ spectra and $\Nch$ distributions. Three-particle cumulants based on the standard and subevent methods are calculated as a function of charged particle multiplicity $\Nch$. To make the results directly comparable to the ATLAS measurement~\cite{Aaboud:2018syf}, the cumulant analysis is carried out using charged particles in $|\eta|<\eta_{\rm{max}}=2.5$ and $0.3<\pT<3$ GeV/c, and the $\Nch$ is defined as the number of charged particles in $|\eta|<2.5$ and $\pT>0.4$ GeV/c.

The $\acn{n,m|n+m}$ is calculated in several steps using charged particles with $|\eta|<2.5$, similar to Refs.~\cite{Jia:2017hbm,Aaboud:2017blb}. Firstly, the correlators $\lr{\{3\}}_{n,m|n+m}$ in Eq.~\ref{eq:1} are calculated for each event from particles in the $\pT$ ranges, $0.3<\pT<3$ GeV/c, and the number of charged particle in this $\pT$ range, $\nchb$, is calculated. Note that $\nchb$ is not the same as $\Nch$ defined earlier due to different $\pT$ ranges used. Secondly, $\lr{\{3\}}_{n,m|n+m}$ are averaged over events with the same $\nchb$ to obtain $\acn{n,m|n+m}$. The $\acn{n,m|n+m}$ values calculated for unit $\nchb$ bin are then combined over broader $\nchb$ ranges of the event ensemble to obtain statistically significant results. Finally, the $\acn{n,m|n+m}$ obtained for a given $\nchb$ are mapped to given $\nch$ to make the results directly comparable to the ATLAS measurements~\cite{Aaboud:2018syf}.

The subevent methods do not necessarily suppress all non-flow contributions. A jet could fall across the boundary between two neighboring subevents. In order to estimate such residual non-flow effects, an additional pseudorapidity gap of 0.5 unit is required between neighboring subevents. The results with and without pseudorapidity gap are compared with each other.

\section{Results and discussion}
 \label{results} 
\begin{figure}[h!]
\begin{center}
\includegraphics[width=0.45\linewidth]{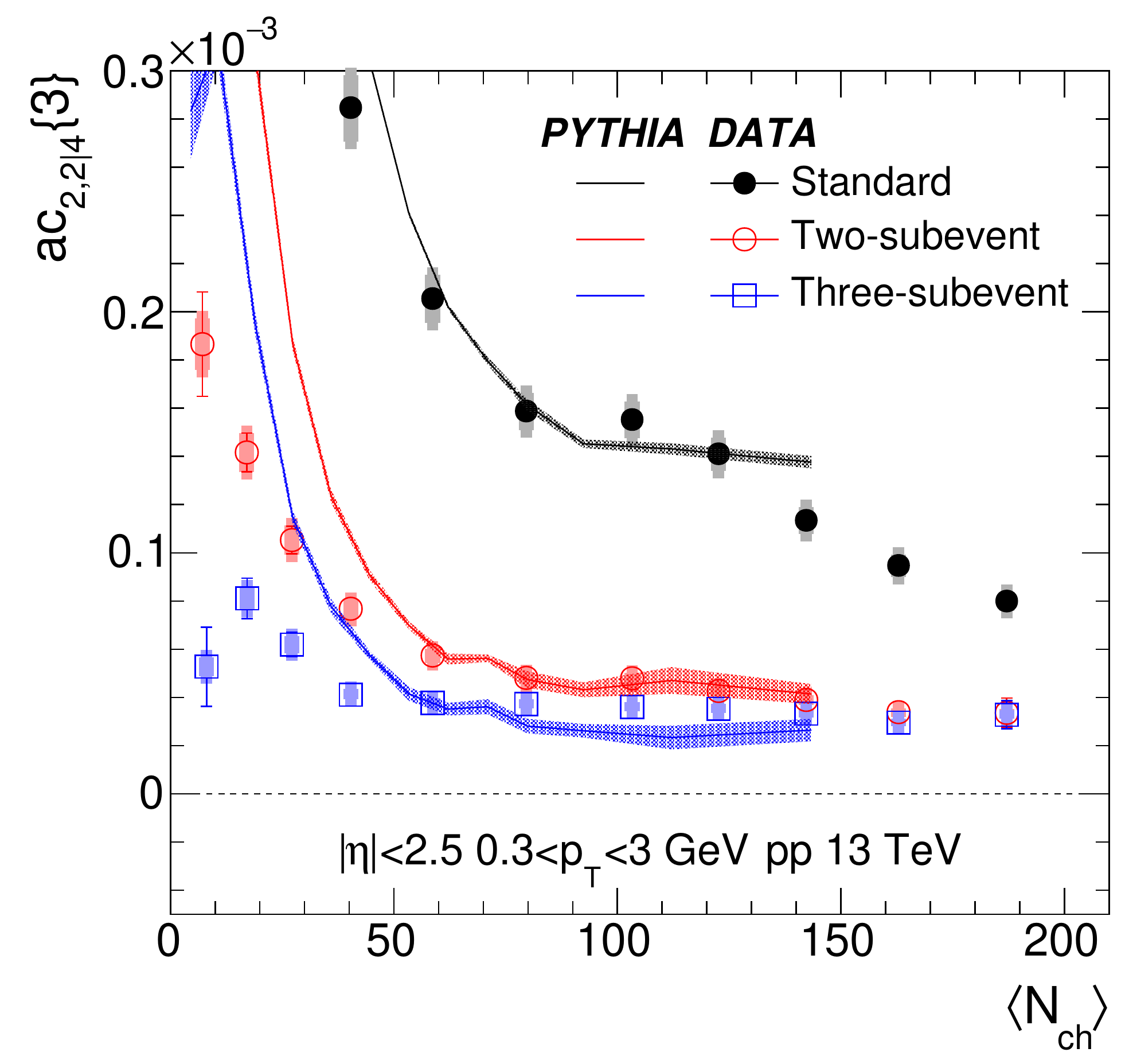}\includegraphics[width=0.45\linewidth]{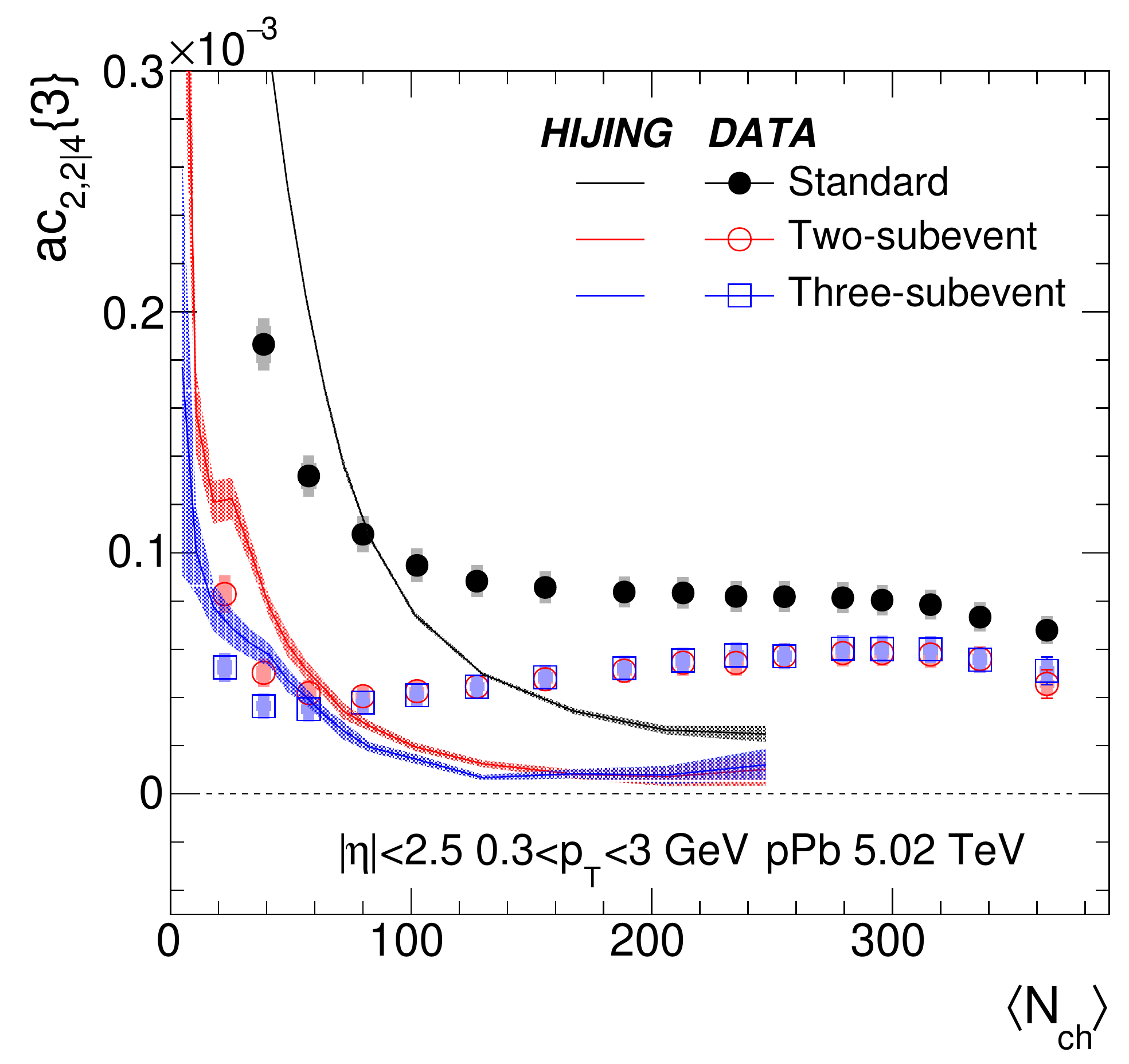}
\end{center}
\vspace*{-0.5cm}
\caption{\label{fig:1} The $\acn{2,2|4}$ calculated for charged particles in 0.3 $<\pT<$ 3.0 GeV/c with the standard, two- and three-subevent cumulant methods as a function of $\nch$ obtained for $pp$ collisions (left panel) and $p$+Pb collisions (right panel). In each panel, the calculations (lines) are compared with the ATLAS data~\cite{Aaboud:2018syf}.}
\end{figure}

Figure~\ref{fig:1} shows the asymmetric cumulant $\acn{2,2|4}$ from the models and compares with the ATLAS $pp$ and $p$+Pb data  for the standard, two- and three-subevent cumulant methods. The $\acn{2,2|4}$ values from standard method are much larger than those from the subevent methods, consistent with the expectation that the standard method is dominated by non-flow contributions from dijets. Significant differences are also observed between the two-subevent and three-subevent methods in $pp$ collisions over the full $\nch$ range and in $p$+Pb collisions for $\nch <$ 150. In $pp$ collisions, the calculated $\acn{2,2|4}$ values decrease sharply up to $\nch \sim $ 60, but decrease very slowly for higher $\nch$. The difference between the two-subevent and three-subevent results are larger than what is observed in the data, suggesting that the non-flow effects are overestimated in PYTHIA8.  In $p$+Pb collisions, $\acn{2,2|4}$ values from HIJING are larger than ATLAS data for $\nch <$ 80, but decrease to below the data for $\nch >$ 80. This implies that the influence of the non-flow effects are subdominant in $p$+Pb collisions at large $\nch$ region, but it still dominates the small $\nch$ region. The results in Figure~\ref{fig:1} suggest that the non-flow correlations are suppressed effectively with the three-subevent method in $p$+Pb collisions, but may potentially still have significant contributions in $pp$ collisions. 

\begin{figure}[h!]
\begin{center}
\includegraphics[width=0.45\linewidth]{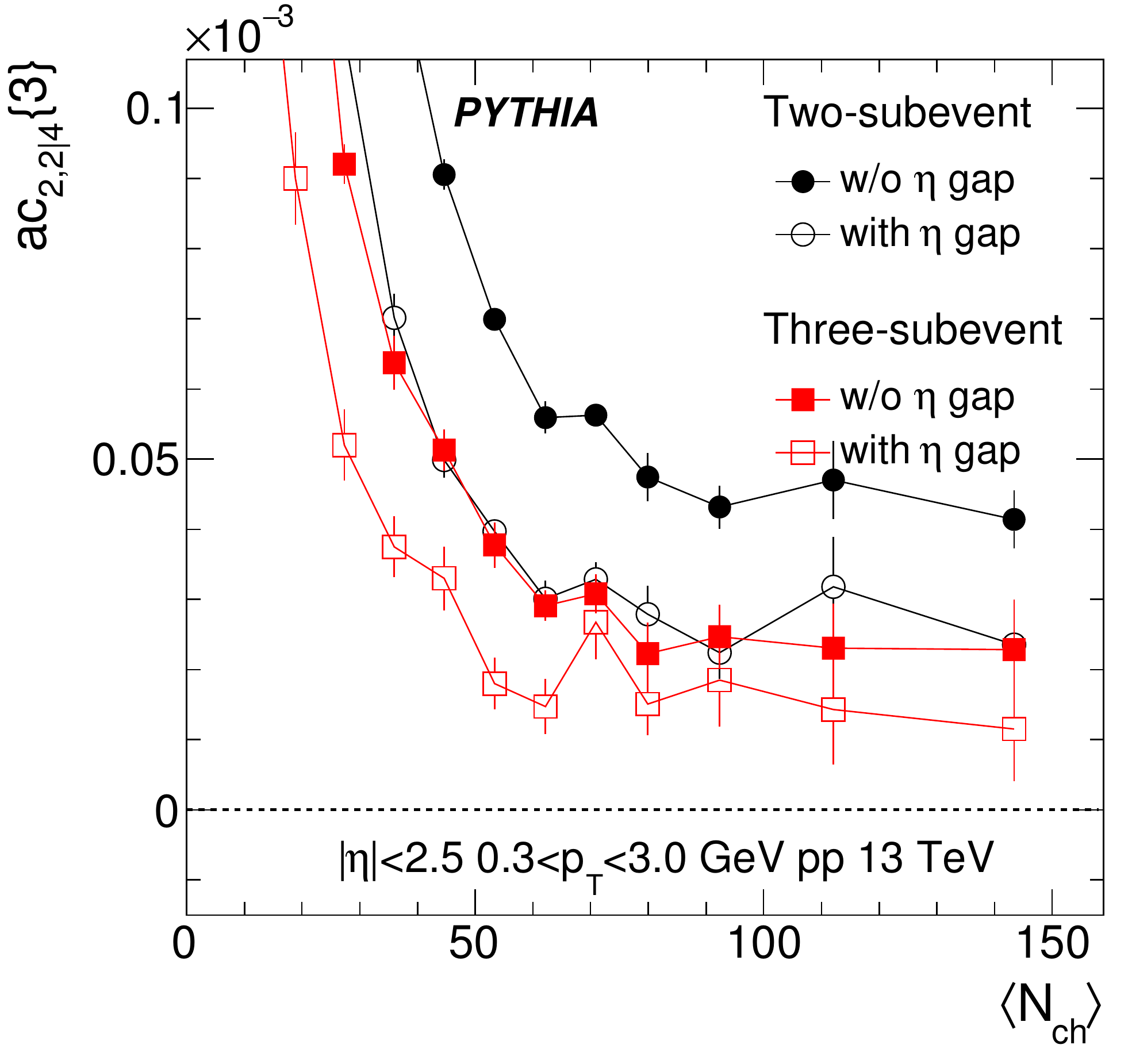}\includegraphics[width=0.45\linewidth]{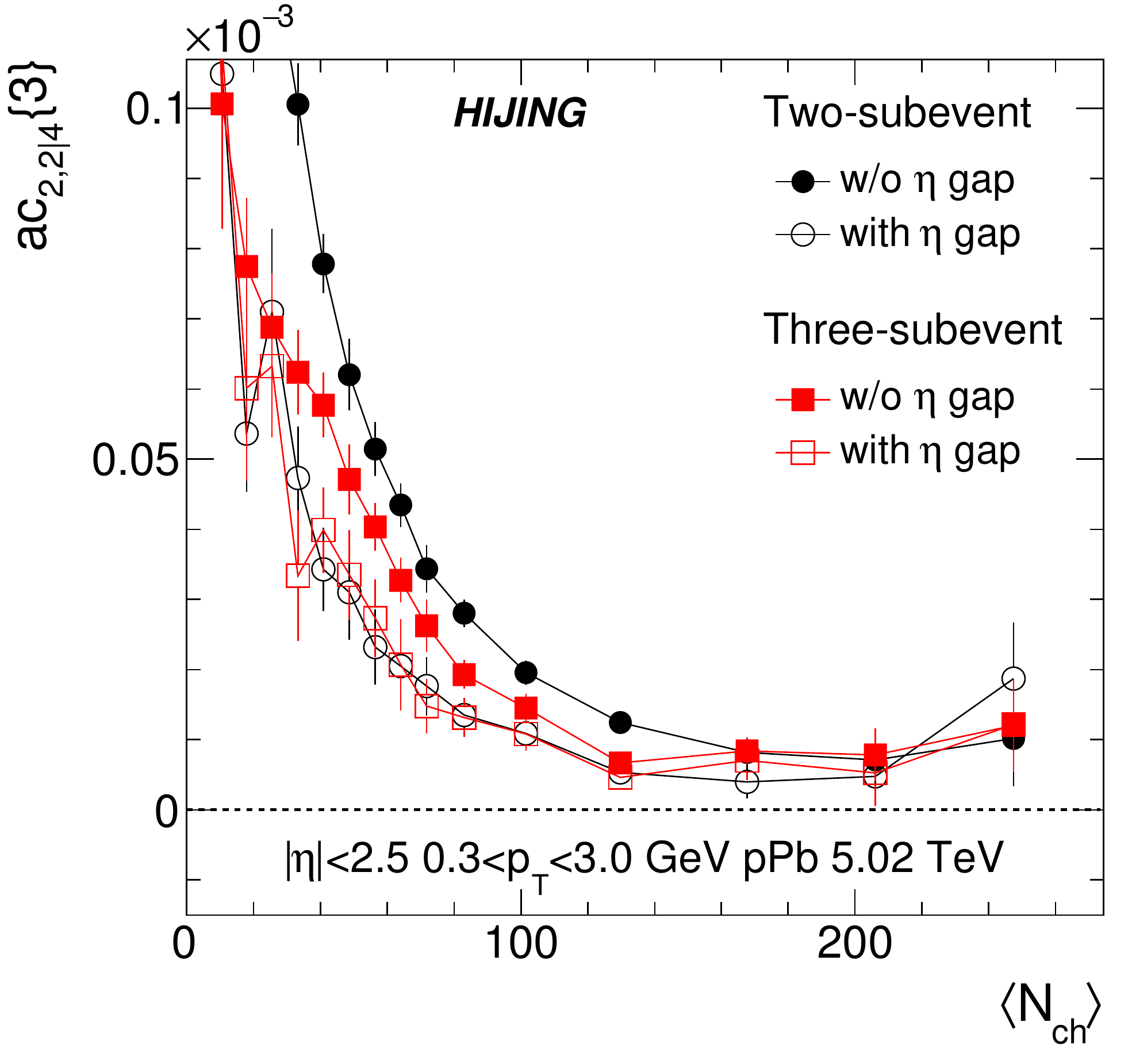}
\end{center}
\vspace*{-0.5cm}
\caption{\label{fig:2} The $\acn{2,2|4}$ calculated for charged particles in 0.3 $<\pT<$ 3.0 GeV/c using the two- and three-subevent cumulant methods with (open symbols) or without (solid symbols) an additional pseudorapidity gap of 0.5 unit between neighboring subevents as a function of $\nch$, for $pp$ collisions (left panel) and $p$+Pb collisions (right panel).}
\end{figure}

To evaluate the effect of a jet falling cross the boundary between two neighboring subevents, we also calculated the $\acn{2,2|4}$ with or without an additional 0.5 unit pseudorapidity gap between neighboring subevents as a function of $\nch$. The results are shown in Figure~\ref{fig:2} for $pp$ and $p$+Pb collisions. The $\acn{2,2|4}$ values were further suppressed with the pseudorapidity gap for both collision systems. It would be very important to repeat the experimental measurement with the same pseudorapidity gap for ALL multi-particle cumulant observables, $c_n\{4\}$, $\scn$ and $\acn{n,m|n+m}$, which shall provide further confidence whether the non-flow effects are under control in the data or not. 

\begin{figure}[h!]
\begin{center}
\includegraphics[width=0.45\linewidth]{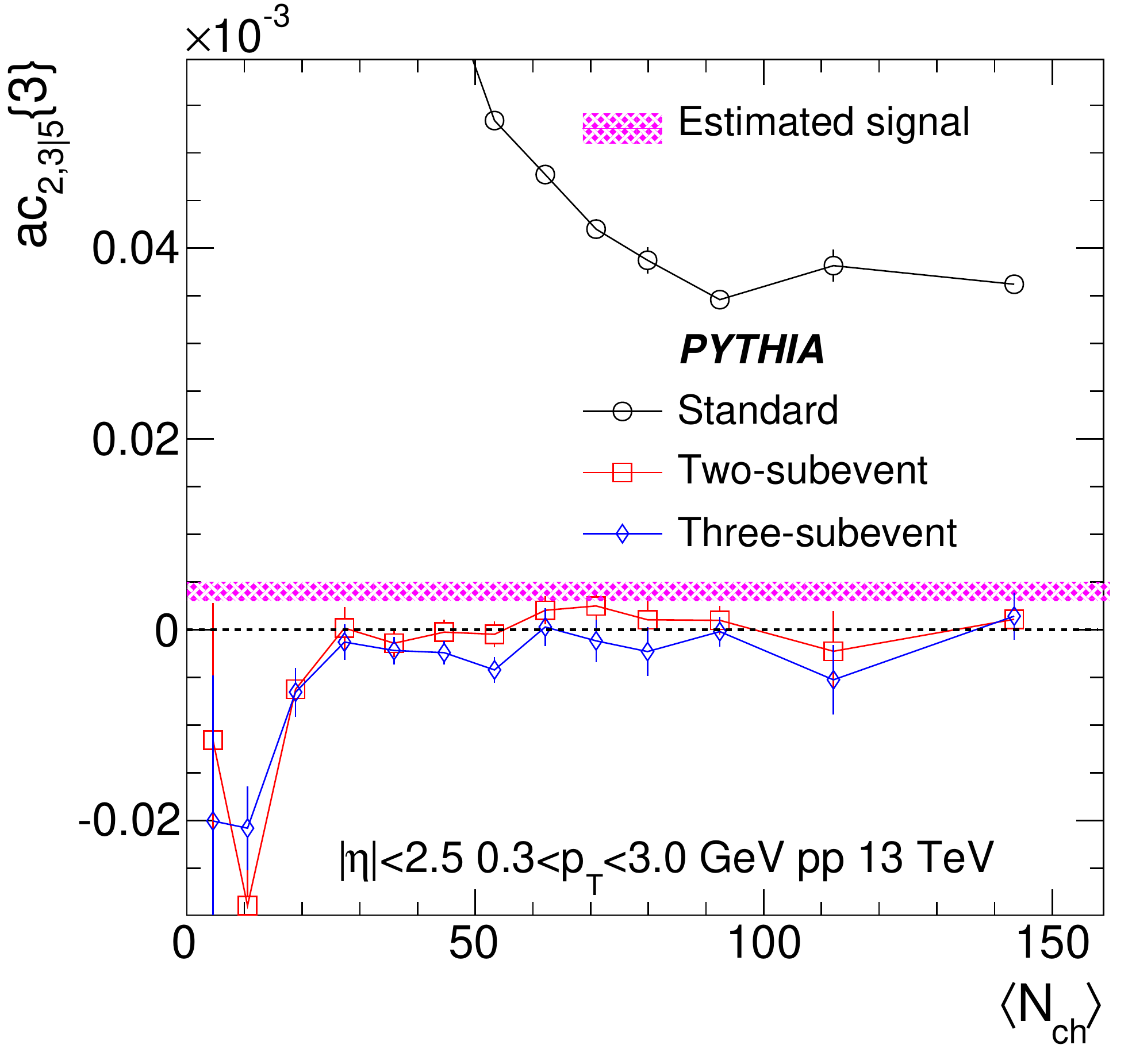}\includegraphics[width=0.45\linewidth]{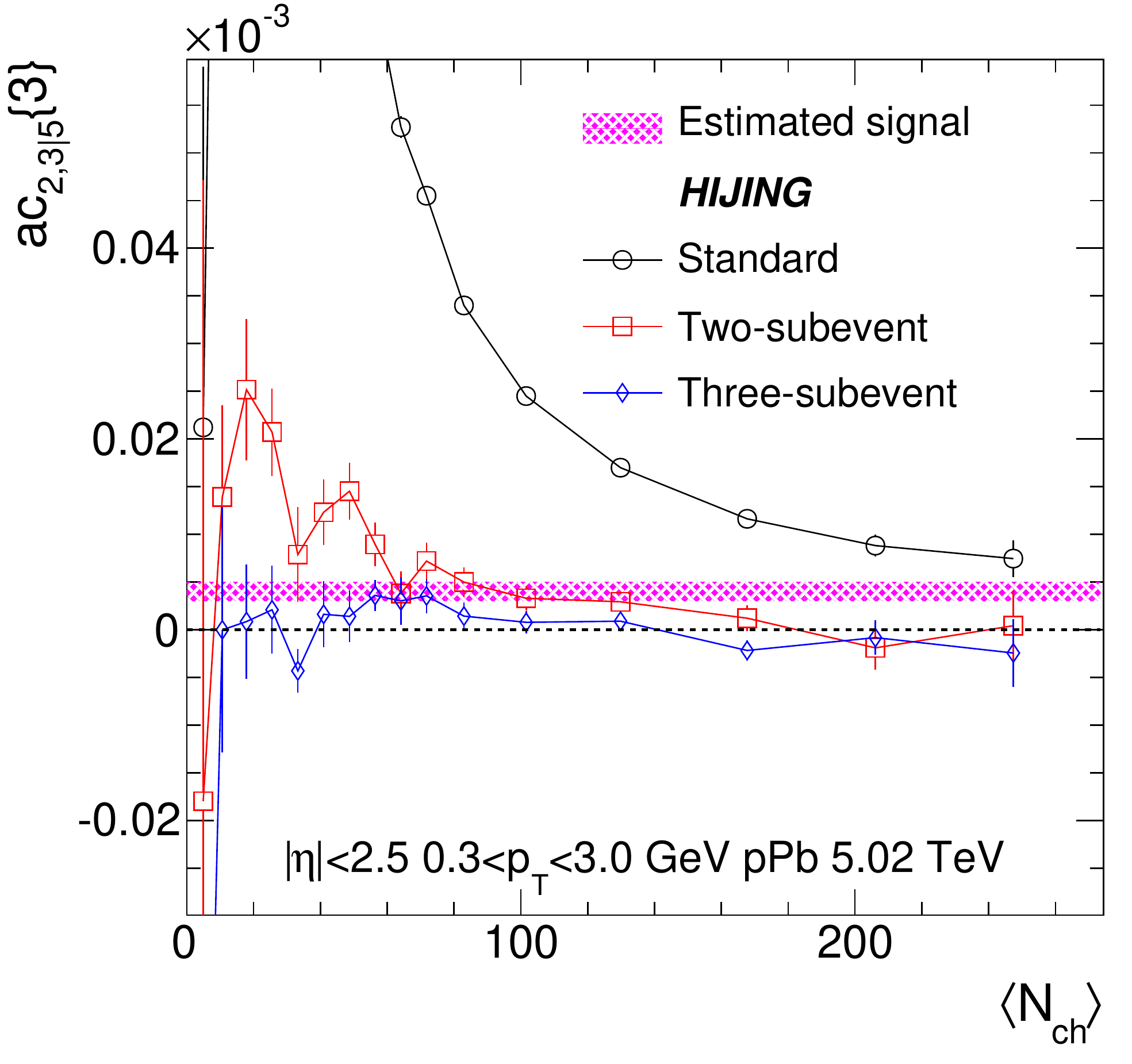}
\end{center}
\vspace*{-0.5cm}
\caption{\label{fig:3} The $\acn{2,3|5}$ calculated for charged particles in 0.3 $<\pT<$ 3.0 GeV/c with the standard, two- and three-subevent cumulant methods as a function of $\nch$ obtained for $pp$ collisions (left panel) and $p$+Pb collisions (right panel). The shaded bands indicate the possible range of the collective signal estimated based on the non-linear response formalism (see text).}
\end{figure}

Figure~\ref{fig:3} shows the prediction of the non-flow effects for the asymmetric cumulant $\acn{2,3|5}$ as a function of $\nch$ for the the standard, two- and three-subevent cumulant methods. The $\acn{2,3|5}$ values from subevent methods are much smaller than those from the standard method. Assuming that the ${\bm V}_5$ is dominated by the non-linear mode coupling effects in peripheral A+A or small collision systems, ${\bm V}_5\approx \chi_{23,5}{\bm V}_2{\bm V}_3$~\cite{Teaney:2012ke,Aad:2015lwa}, one expects that $\acn{2,3|5}\approx\chi_{23,5}\lr{v_2^2v_3^2}$, where $\chi_{23,5}$ is the non-linear response coefficients~\cite{Yan:2015jma}. The value of $\chi_{23,5}$ is measured to be $\chi_{23,5}\approx 2-3$ in Pb+Pb collisions, and is nearly constant toward peripheral collisions~\cite{Acharya:2017zfg}. The $\chi_{23,5}$ is not yet measured in $pp$ and $p$+Pb collisions, and we shall assume that it is the same as in Pb+Pb collisions. Based on the measured $v_n$ in $pp$ and $p$+Pb collisions, i.e., $v_2\approx 0.06$ and $v_3 \approx 0.02$~\cite{Aaboud:2018syf}, we estimate the $\acn{2,3|5}$ signal associated with the long-range collectivity is about 3 $\sim$ 5$\times 10^{-6}$. This signal, shown as a shaded band in Figure~\ref{fig:3}, is comparable or slightly larger than the residual non-flow in the three-subevent cumulant, and therefore should be measurable in the high-multiplicity region of $pp$ and $p$+Pb collisions. Note that the values of $\acn{2,3|5}$ have a tendency to decrease and become negative in small $\nch$ region. The origin of this is related to the anti-correlation between $v_2$ and $v_3$ caused by inter-jet correlations.~\footnote{Correlations associated with the away-side jet tend to increase $v_2^2$ and decrease the $v_3^2$, eventually lead to a negative $v_3^2$ values in the very low $\nch$ region. Such anticorrelation between $v_2$ and $v_3$ is also responsible for the large negative symmetric cumulant $\sca$ values in the low $\nch$ region of $pp$ and $p$+Pb collisions~\cite{Huo:2017nms}.}

In summary, we calculated the three-particle mixed-harmonic correlator $\acn{2,2|4}=\lr{v_2^2v_{4}\cos4(\Phi_2-\Phi_{4})}$ and $\acn{2,3|5}=\lr{v_2v_3v_{5}\cos(2\Phi_2+3\Phi_3-5\Phi_{5})}$ in $pp$ and $p$+Pb collisions from PYTHIA8 and HIJING models. These models do not have the genuine long-range collectivity, and therefore provide estimation for the possible contributions from non-flow effects. We show that the three-subevent methods can significantly suppress the non-flow from jets and dijets, as argued by the ATLAS measurement. For $\acn{2,2|4}$, the residual non-flow effects are much smaller than the measured collectivity signal in $p$+Pb collisions for $\nch>80$, but could still be important in $pp$ collisions. For $\acn{2,3|5}$, the residual non-flow effects are comparable to or smaller than the estimated signal based on the non-linear response formalism, therefore this correlator should be detectable in LHC experiments. To further suppress these non-flow effects, experimental measurements should be repeated with a requirement of pseudorapidity gap between subevents. The subevent method can also be generalized to other mixed-harmonic correlators with three or more particles, such as $\lr{{\bm V}_2{\bm V}_{4}{\bm V}_{6}^*}$ and $\lr{{\bm V}_2{\bm V}_{2}{\bm V}_{2}{\bm V}_{6}^*}$, although the relative importantance of collectivity and non-flow deserves more detailed study.

We thank Mingliang Zhou for generating the PYTHIA events. This work is supported by National Science Foundation grant number PHY-1613294, MOST of China under 973 Grant 2015CB856904, the National Science Foundation of China No. 11475243, 11421505 and A050306, and Chinese Scholarship Council No. 201704910762.

\bibliography{asc}{}
\bibliographystyle{apsrev4-1}

\end{document}